\documentclass[12pt,a4paper,final]{iopart}
\pdfoutput=1
\usepackage{iopams}  
\usepackage{graphicx}
\usepackage[colorlinks=true,linkcolor=blue,urlcolor=blue,citecolor=blue]{hyperref}

\usepackage[
backend=biber,
style=numeric,
sorting=ynt,
citestyle=numeric-comp
]{biblatex}
\usepackage{comment}
\usepackage{xstring}
\usepackage{bbold}
\usepackage[nottoc]{tocbibind}

\begin{document}

\title[Swarm parameter measurement in hydrogen, considering secondary photonic electron emission]{Swarm parameter measurement of hydrogen, considering secondary photonic electron emission}

\author{Andreas H\"osl}


\author[cor1]{Christian Franck}
\address{ETH Z\"urich, High Voltage Lab}
\eads{\mailto{franck@eeh.ee.ethz.ch}}

\begin{abstract}
Discharges in hydrogen at pressures above $1\,$kPa and a reduced electric field of $E/N=100-200\,$Td show a characteristic current oscillation in Pulsed Townsend experiments. This is explained by secondary emission of electrons from the photo-cathode: some hundreds of nano-seconds after the laser-pulse that released the initial $10^4-10^6$ primary electrons, secondary electrons are emitted from the cathode. Mechanisms discussed in literature are UV-emission from neutral molecules, emission by positive ions reaching the cathode, and back-scattering of excited neutrals. For a measurement up to $500\,$Td and pressures up to $2\,$kPa we model different sources, and agree with previous findings that the observed secondary emission is purely due to ultra-violet light below $200\,$Td: the simulation fits with the complicated form of the oscillating waveform. Obtained swarm parameters agree well with the literature. Our findings suggest a very high efficiency of the photo-cathode for UV light of energies above $8\,$eV.
\end{abstract}


\section{Introduction}
\begin{figure}[htb]
    \centering
    \includegraphics[width=0.8\textwidth]{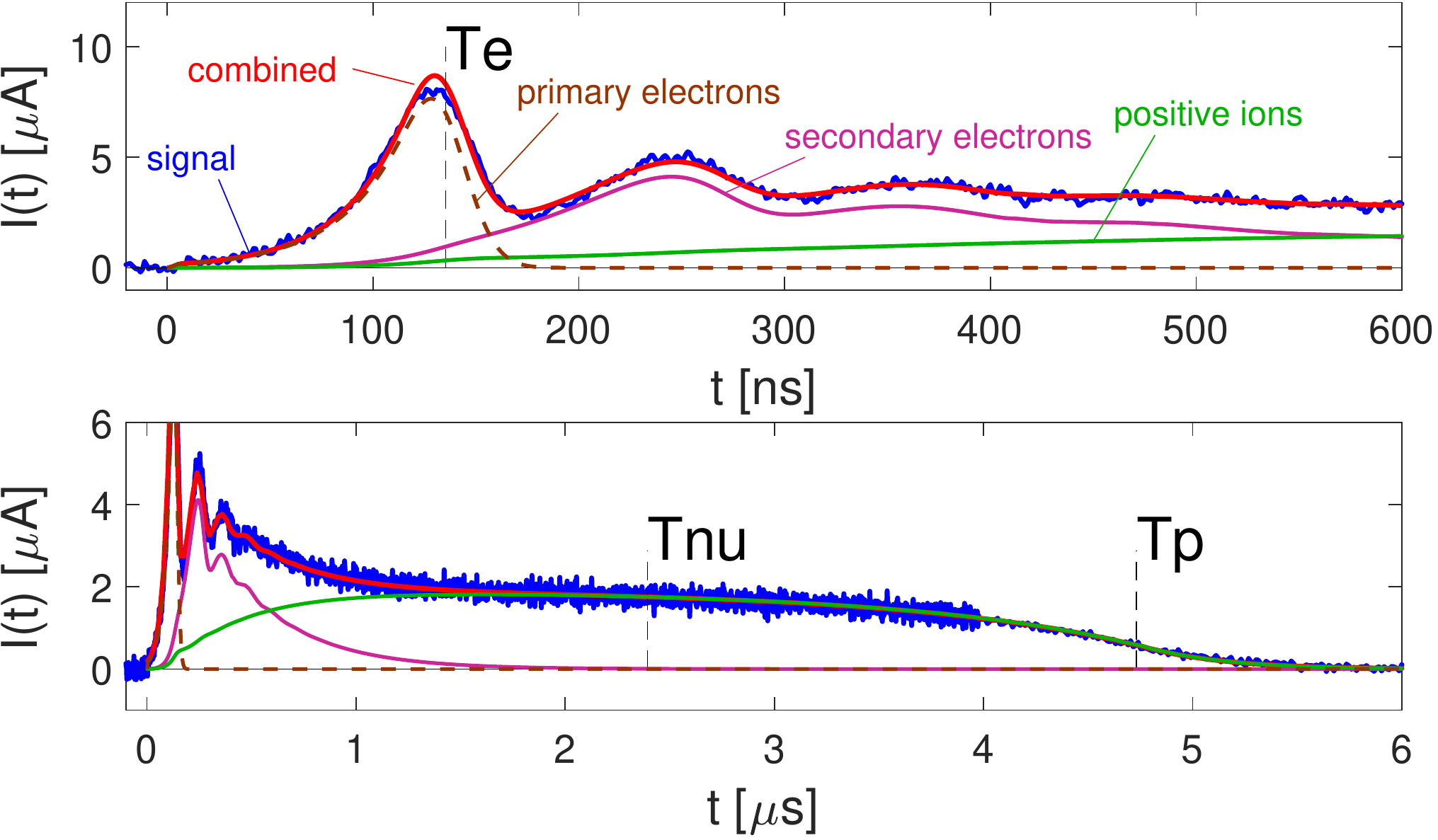}
    \caption{A measurement at $130\,$Td, at a pressure of $1.1\,$kPa and $25\,$mm gap distance, on two different time scales. For times before the electron drift time $T_{\mathrm{e}}=130\,$ns, the typical exponential growth of electrons is observed. Secondary peaks are found at (roughly, but not exactly) multiples of $T_{\mathrm{e}}$. The charge contribution of the secondary electrons outweighs the primary electrons by far. The current after $1\,\mu$s is then mainly positive ions.}
    \label{fig:waveform}
\end{figure}
The well established Pulsed Townsend method allows the measurement of swarm parameters in a variety of gases: a pulsed UV laser releases electrons from a photo-cathode, which then travel in the applied homogeneous electric field and interact with neutral gas molecules. The displacement current of the charge carriers is measured and evaluated.\\
The discharge in hydrogen at few kPa pressure and $E/N=100-200\,$Td shows an unusual, oscillating temporal development. A release of electrons some time after the initial laser pulse, and by a different source than the laser, has to be taken into account in order to evaluate the measured waveforms ("secondary emission"). Figure \ref{fig:waveform} shows an example of such a measurement.\\
Our motivation for examining hydrogen discharges is mainly the investigation of the photo-cathode. We use hydrogen to "regenerate" the cathode after measuring oxidizing or fluoridating gases, which tend to decrease the efficiency (released electrons per UV photon). After being subjected to a hydrogen discharge at $200-400\,$Td or higher, and at pressures from $50-300\,$Pa, its efficiency typically increases by a factor of $2-10$. The reason for this is not entirely clear.
Secondly, we aim at modeling secondary emission in order to be able to distinguish between secondary emission and ion detachment effects in other gases, which exhibit large currents at ion-timescales.\\
The physics of hydrogen can become very complicated at high reduced electric fields above $500\,$Td; an extensive overview of which is given by Phelps et al ~\cite{phelps1993oscillations,phelps2009energetic}. We consider and implement a subset of the variety of discussed cathode phenomena, which are potentially relevant below $500\,$Td:
\begin{itemize}
    \item Excitation of the photo-cathode due to positive ions or neutral excited molecules: positive ions neutralize upon arrival at the (photo-)cathode; if the recombination energy is sufficient to remove two electrons (one to neutralize, one to emit) from the cathode material, secondary emission could energetically be possible. The cathode materials are specifically chosen for their low work function, which is usually around $4-5\,$eV, whereas the the ionization energy of hydrogen is $15.4\,$eV. Phelps~\cite{phelps2009energetic} gives a different physical explanation and assumes that fast neutral $H$ atoms (equation~\ref{eq:H}) are responsible for secondary emission, as observed by Fletcher and Blevin~\cite{fletcher1981nature} above $220\,$Td.
    \item Secondary emission due to UV-photons, for which we consider two variants: First, excitation of neutral molecules upon collision with free electrons, followed by radial emission of a UV photon. Secondly, UV emission upon arrival of an electron at the anode.
\end{itemize}
Phelps~\cite{phelps2009energetic} agrees with Blevin and Fletcher~\cite{fletcher1981nature}, that below $300\,$Td, photo-emission is the dominating process for secondary emission. Our findings further confirm this: modeling UV emission by neutral molecules, proportional to the spatial distribution of electrons, reproduces all waveforms up to $200\,$Td, and fails above.
Phelps and other sources~\cite{phelps2009energetic,stefanovic1997volt} state further that positive ions are the primary source of secondary electron generation above $250\,$Td. We try a simplified implementation thereof, yet fail to achieve a good overlap of measurement and simulation.\\
The evaluation of the waveforms is done using a finite volume simulation on GPUs, as is described in~\cite{hosl2017measurement}.\\


\section{Kinetic Model}
\label{ch:kinMod}
\subsection{Physical picture}
Molecular hydrogen has, compared to other non-noble gases, a comparably high ionization threshold of $15.4\,$eV:
\begin{equation}
    \label{eq:ionization}
    e + H_2 \stackrel{\;\nu_{\mathrm{i}}\;}{\rightarrow} 2e + H_2^{+}.
\end{equation}
We model this with a constant rate $\nu_{\mathrm{i}}$, and assume that the electron energies are in thermodynamic equilibrium. At high $E/N$ and low pressures, this introduces a certain error, as soon as ionization time-scales are of the same magnitude as equilibration. The error is difficult to estimate, and was discussed for instance in~\cite{korolov2016scanning}.\\
The ion clustering
\begin{equation}
    \label{eq:chargetransfer}
    H^{+}_2 + H_2 \rightarrow H_3^{+} + H,
\end{equation}
is very efficient~\cite{dawson1962detection}, and indeed we find ion mobilities that are compatible with reference values for $H_3^{+}$.
Dissociative ionization is considered in the complex model of Phelps et al~\cite{phelps2009energetic}, and accounts for a few percent of the total ionization.\\
For $H_{3}^{+}$ ions of sufficient energy a dissociation into $H^{+}$ and $H_{2}$ is observed above $250\,$Td. Phelps states that subsequent charge transfer
\begin{equation}
\label{eq:H}
H_{2}+H^{+}\rightarrow H_{2}^{+}+H
\end{equation}
would then create fast neutral $H$ atoms which are capable of releasing electrons from the photo-cathode. Another possible source is subsequent chemo-ionization and associative ionization
\begin{equation}
\label{eq:H3}
    H_{2}^{*} + H_2 \rightarrow H_{3}^{+} + H + e
\end{equation}
\begin{equation}
    H^{*} + H_2 \rightarrow H_{3}^{+} + e
\end{equation}
with thresholds of $13.7\,$ and $14.7\,$eV, as considered in~\cite{dehmer1995rydberg}.\\

\noindent The $H_2$ dissociative attachment rate
\begin{equation}
    \label{eq:attachment}
    e + H_2 \stackrel{\;\nu_{\mathrm{a}}\;\;}{\rightarrow} H^{-} + H,
\end{equation}
is $2-3$ orders of magnitudes smaller than the ionization rate over the whole measurement range. The energy required is $7\,$eV according to Biagi's cross section set~\cite{biagi1997magboltz} from MAGBOLTZ (version 8.9, source: LXCat~\cite{lxcat}), which differs from values of the TRINITY cross section set~\cite{dyatko2011eedf}, also on LXCat, which features values around $3.6\,$eV.
Detachment of the electron of $H^{-}$ has, for instance, been investigated in~\cite{huq1983electron}.
We are unable to observe neither attachment nor detachment in our measurements, as its current contribution is by orders of magnitude lower compared to positive ions and primary electrons. We thus use the results of a Bolsig+ simulation to preset the attachment rate coefficient.
Omitting this rate, or introducing detachment, does not visibly change the current shape of our measurements.\\

\noindent The measured total displacement current of the Pulsed Townsend experiment is then given as
\begin{equation}
    I_{\mathrm{tot}}(t) = I_{\mathrm{e}}(t) + I_{\mathrm{p}}(t) + I_{\mathrm{n}}(t)=
    \frac{q_0}{d} \int_{0}^{d} \frac{\rho_{\mathrm{e}}(x,t)}{T_{\mathrm{e}}} + \frac{\rho_{\mathrm{p}}(x,t)}{T_{\mathrm{p}}} + \frac{\rho_{\mathrm{n}}(x,t)}{T_{\mathrm{n}}} \mathrm{d}x,
\end{equation}
with elemental charge $q_0$, densities $\rho(x,t)$ of electrons (e), positive ions (p), and negative ions (n), gap drift times $T_{e}$, $T_{p}$ and $T_{n}$. The integral is taken over the gap distance $d$.\\

\subsection{Fitting method}
We use our established method of fitting simulations to measurements, as described in~\cite{hosl2017measurement}. The finite-volume simulations run on GPUs, to cope with the large computational cost. In the following we describe the additional extensions in order to simulate secondary emission.

\subsection{Secondary emission modeling}
\begin{itemize}
    \item Model 1, secondary emission due to arriving positive ions/neutral excited atoms at the photo-cathode is modeled straight-forward: at a certain probability, a new electron is released per arriving positive ion.
    \item Model 2.1, Proportional to the electron density $\rho_{\mathrm{e}}(x,t)$ at position $x$, excited neutral molecules are created. If the life-time of the excitation is below few nano-seconds, the decay of this excitation can be modelled as instantaneously, and is followed by a radial emission of a UV photon (equation (\ref{eq:S11})).
    \item Model 2.2, UV light is emitted upon arrival of an electron at the anode, which releases an energy equal to the kinetic energy of the electron plus, possibly, the "work function" of the anode material. This is implemented similar to model 1.
\end{itemize}

\noindent The equation describing model 2.1 is given as
\begin{equation}
\label{eq:S11}
    \frac{\partial}{\partial t} \rho_{\mathrm{e}}(x=0,t) = \ldots + 
    \nu_{\mathrm{UV}}\,\int_{0}^{d} \rho_{\mathrm{e}}(x,t) \, g(x) \,\mathrm{d}x.
\end{equation}
Proportional to the electron density $\rho_{\mathrm{e}}(x,t)$ at position $x$, UV light is emitted and releases electrons from the photo-cathode at $x=0$. Absorption of UV light in the gas is assumed to be negligible. $g(x)$ is a geometric factor, i.e. probability to hit the photo-cathode for radially emitted light. Our calculation of $g(x)$ assumes an even distribution of electrons over our cathode of $1.25\,$cm radius, and neglects transversal diffusion. The geometric factor is maximal at $x=0$ at $50\,$\%, and drops to roughly $10-20\,$\% at the anode for the measurement gap distances of $1.5-3\,$cm.\\


\section{Results}
\label{ch:results}
Trying out the different variants of secondary emission, we find that
\begin{itemize}
    \item Model 1, in which positive ions/neutral atoms release electrons at impact at the cathode does not recreate the measured waveforms. Neither does a combination with UV emission (model 2.1), which we try out for the $E/N$ range beyond $250\,$Td.
    \item Model 2.1, instantaneous UV-emission, explains all waveforms up to $200\,$Td. Since the temporal evolution of the waveforms as shown in the example plot~\ref{fig:waveform} is relatively complex, we are lead to believe that this is the correct model.
    \item Model 2.2, in which UV is emitted upon arrival of the electrons of the anode reproduces an oscillations, yet the secondary maxima positions do not fit with the measurement. The model is periodic with the electron drift time, whereas the measured waveforms feature maxima slightly earlier than $2\,T_{\mathrm{e}}, \,3\,T_{\mathrm{e}},\,\ldots$.
\end{itemize}
\begin{figure}[ht]
    \centering
    \includegraphics[width=0.8\textwidth]{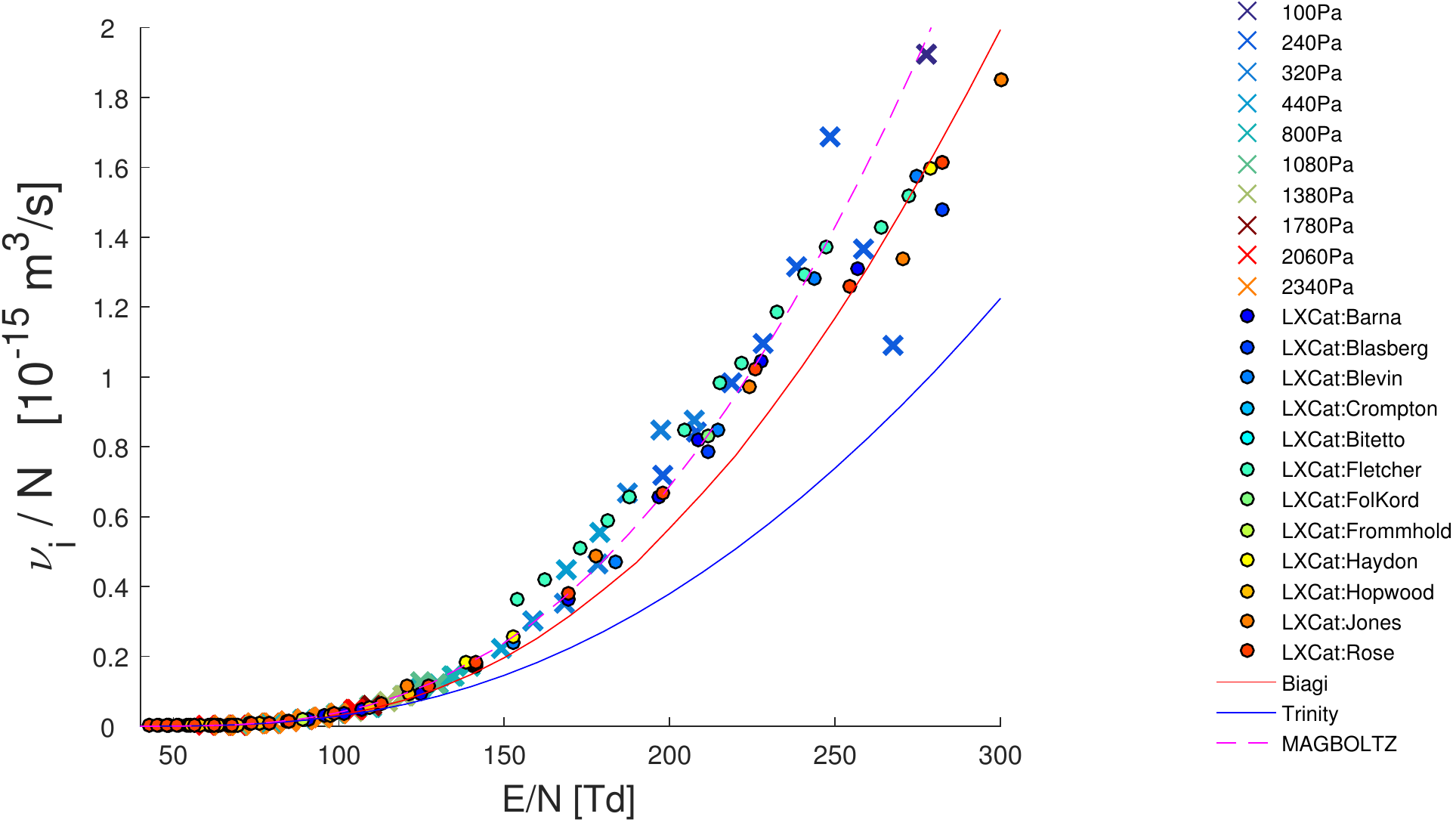}
    \caption{The fitted ionization rate coefficient for different measured pressures and $E/N$. Our values are compared to literature values from LXCat~\cite{lxcat} by different authors, as well as Bolsig+ simulations using the Biagi~\cite{biagi1997magboltz} and Trinity~\cite{dyatko2011eedf} database.}
    \label{fig:nu_i}
\end{figure}
As an output of the fitting process, we obtain the following rate coefficients for model 2.1, assuming instantaneous UV emission of neutral molecules. 
We compare our findings to literature results of various authors on LXCat~\cite{lxcat}, as well as Bolsig+ simulations (2-term Boltzmann approximation) using Biagi's data set (from MAGBOLTZ code version 8.9~\cite{lxcat}) and the TRINITY data set~\cite{dyatko2011eedf}.
We further compare to a MAGBOLTZ simulation (Monte-Carlo) with a newer data set, version 11.2~\cite{biagi1997magboltz}.\\
All references agree with our findings for the ionization rate coefficient below $250\,$Td, as shown in figure \ref{fig:nu_i}. Above, our values show a large spread, while the quality of the fit degrades strongly: currents after the gap crossing time of the positive ion begin to emerge, which cannot be explained within the model.
Therefore, our findings at high $E/N$ are should not be regarded as "fitting results", but should rather illustrate the incipient failure of our model and hint at a different channel of secondary emission.\\
\begin{figure}[ht]
    \centering
    \includegraphics[width=0.8\textwidth]{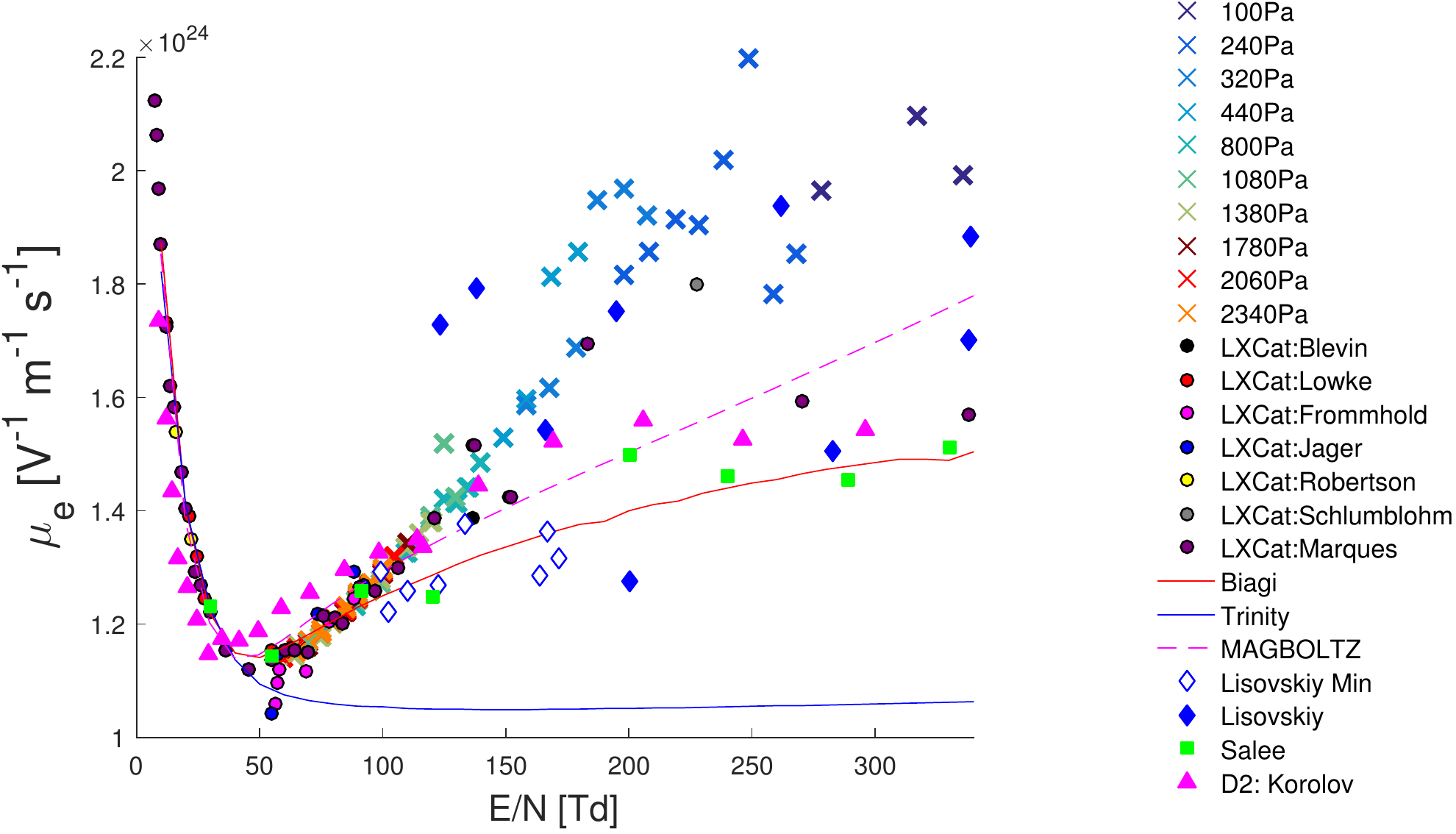}
    \caption{The electron mobility, defined as $w_{\mathrm{e}} = \mu_{\mathrm{e}}\,\cdot \,E/N$, is plotted against $E/N$, and compared to reference data and Bolsig+ simulations~\cite{biagi1997magboltz,dyatko2011eedf,lxcat,lisovskiy2006electron,saelee1976}; furthermore data for deuterium~\cite{korolov2016scanning}.}
    \label{fig:e_mobility}
\end{figure}\\
As for the ionization rate coefficient, our values for the electron mobility, figure \ref{fig:e_mobility}, agree with reference values below $150\,$Td, and deviate for higher $E/N$, where most referenced authors find a lower mobility. It is not unlikely that our values there are fitted too high: at the lower measurement pressures, we are unable to observe the arrival of the electrons directly, due to strong longitudinal diffusion; therefore the fitting is rather indirect.\\
\begin{figure}[ht]
    \centering
    \includegraphics[width=0.8\textwidth]{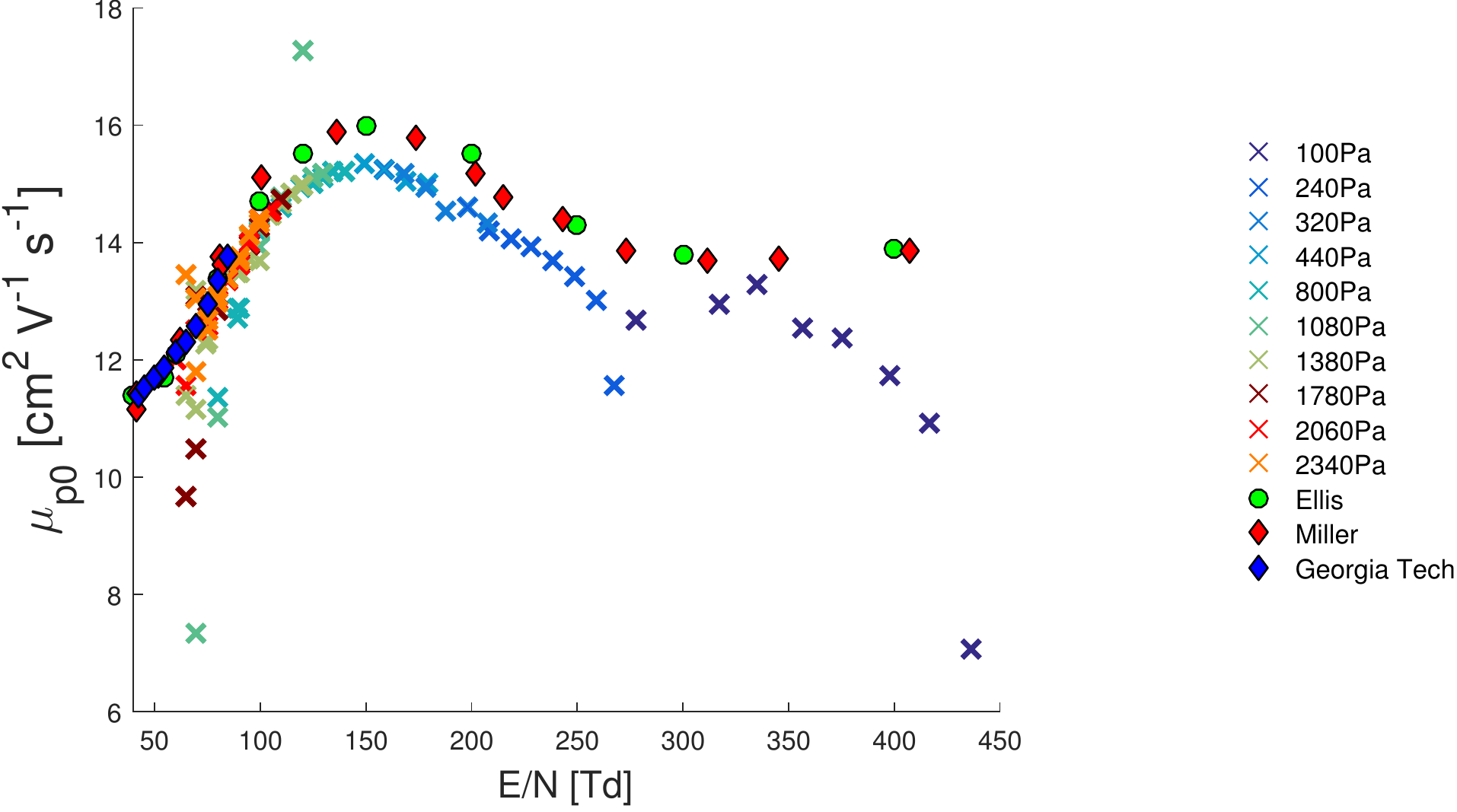}
    \caption{The fitted ion mobility for the positive ion is compared to reference data for $H_{3}^{+}$~\cite{ellis1976transport,miller1968reactions,lxcat} from LXCat. The mobility is normalized to particle density $N=2.686 \cdot 10^{25}\, \mathrm{m}^{-3}$ ("standard conditions"), in order to ensure comparability to older literature values. The measurement was done at a temperature of $25^{\circ}$C.}
    \label{fig:p_mobility}
\end{figure}\\
Figure \ref{fig:p_mobility} shows the fitted mobility of the positive ion, which is presumably mainly $H_{3}^{+}$. Below $100\,$Td, our measurement is not suitable for deducing this mobility since the ionization is low and the ion signal too weak, resulting in a strong spread. We find lower values than Miller et al~\cite{miller1968reactions}, Ellis et al~\cite{ellis1976transport}, and reports from Georgia Tech (1970-74) (the latter two taken from the Viehland database~\cite{viehland1995relating} on LXCat~\cite{lxcat}) over the whole measurement range.\\
\begin{figure}[ht]
    \centering
    \includegraphics[width=0.8\textwidth]{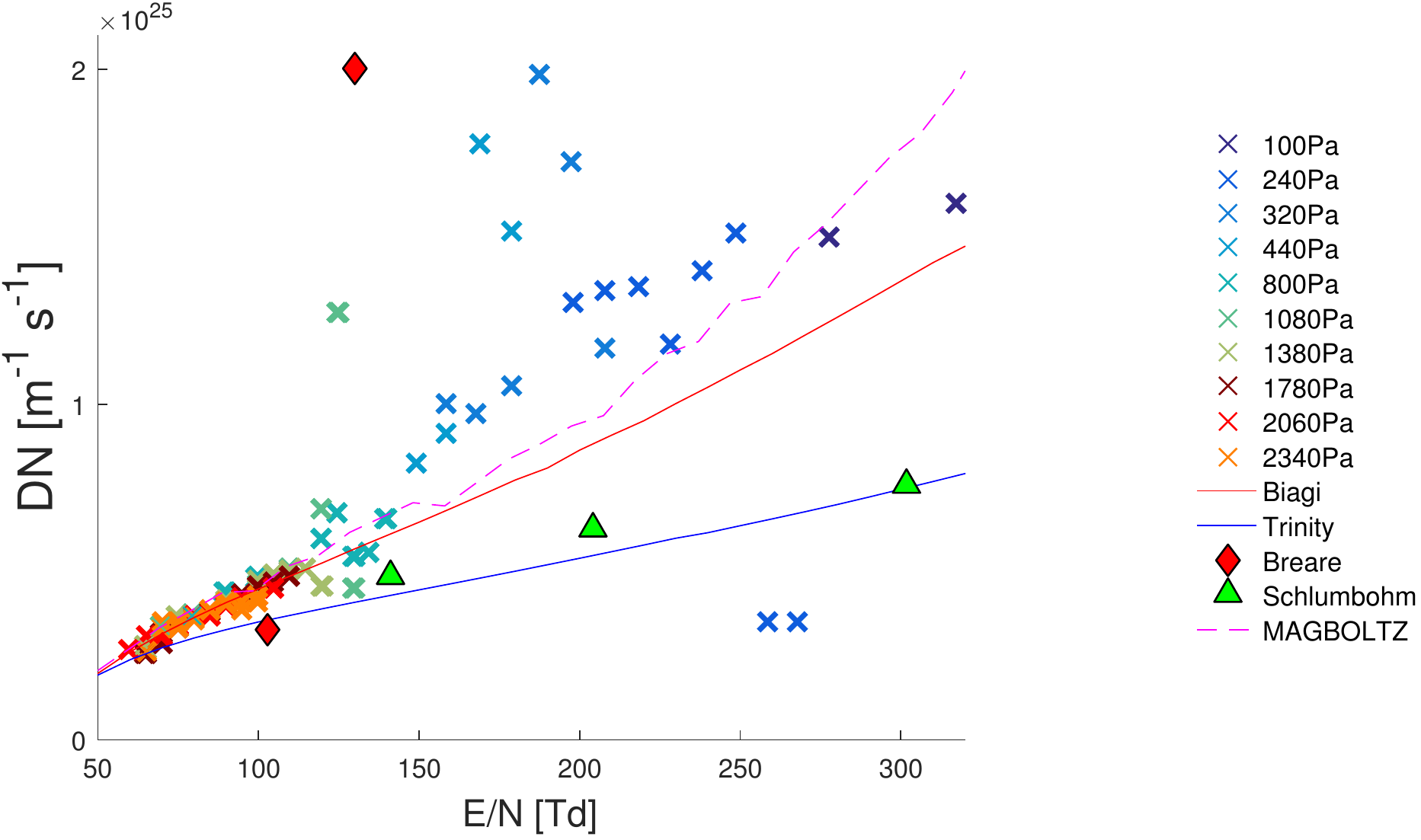}
    \caption{The fitted diffusion is compared to reference values of ~\cite{breare1964locating,schlumbohm1965messung} as well as Bolsig+ simulations using Biagi's dataset (MAGBOLTZ version 8.9)~\cite{lxcat}, TRINITY database~\cite{dyatko2011eedf}, and a MAGBOLTZ simulation (version 11.2)~\cite{biagi1997magboltz}.}
    \label{fig:diffusion}
\end{figure}\\
The density-normalized diffusion of hydrogen is shown in figure \ref{fig:diffusion}. Our values seem to support the MAGBOLTZ simulation and the Bolsig+ calculations based on Biagi's dataset. Above $150\,$Td, the quality of the fit degrades strongly.\\
\begin{figure}[ht]
    \centering
    \includegraphics[width=0.8\textwidth]{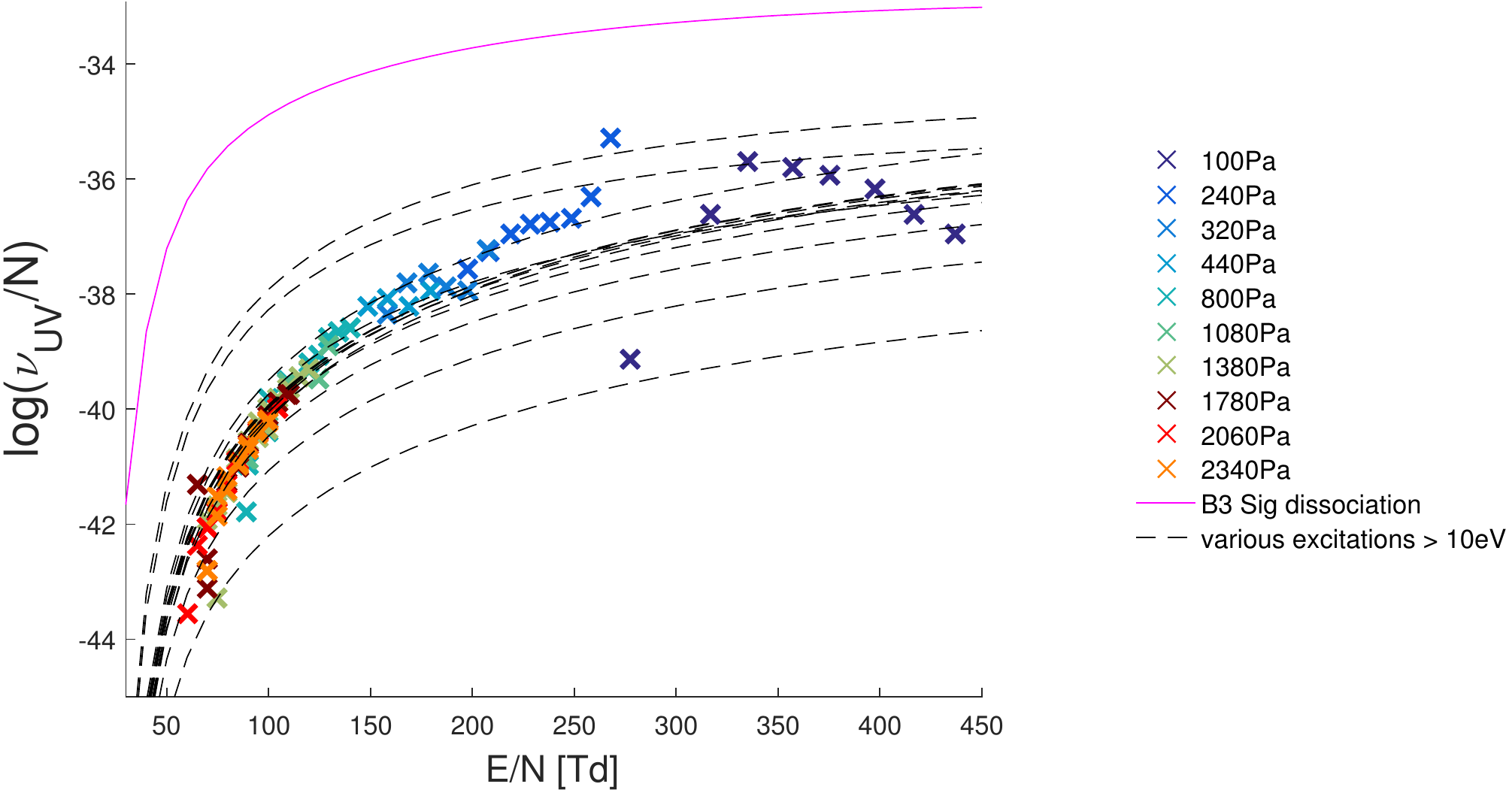}
    \caption{The fit for the UV secondary emission shows an exponential increase with increasing $E/N$. For comparison, excitation rate coefficients of $H_2$ are plotted according to a Bolsig+ simulation using Biagi's database~\cite{biagi1997magboltz}.}
    \label{fig:UV_reionization}
\end{figure}\\
In order to explain the measured current, a UV light emission rate coefficient was introduced in the model. We fit an (over-)exponential increase (figure \ref{fig:UV_reionization}). Surprisingly, the various excitations of hydrogen (according to Bolsig+ using Biagi's cross sections) seem to be of the same magnitude as the fitted UV emission rate. Furthermore, there is no excitation of similar photon energy as the laser which we use for back-illumination, but rather at energies of $8\,$eV and more. 

\section{Discussion}
Using a simple model for secondary emission, we are able to fit measurements up to $200\,$Td and recreate the complex waveform of the hydrogen discharge in homogeneous fields for pressures up to $2\,$kPa. Our results for the swarm parameters fit well to literature, and serve as a check that this model works well. 
We therefore agree with Phelps~\cite{phelps1993oscillations,phelps2009energetic}, and Fletcher and Blevin~\cite{fletcher1981nature}, that photonic secondary electron emission dominates below $200\,$Td, and present for the first time an evaluation in a Pulsed Townsend experiment of the oscillating current. We are, however, unable to fit the waveforms at higher $E/N$.\\
It is difficult to decide which of the hydrogen's excitations is responsible for the secondary emission without measuring the emitted spectrum. The supposed UV emission rate implies a very high efficiency of the electron release: Summing up every possible excitation of the Bolsig+ simulation yields a rate that is only a factor of $50-100$ larger than what is required to explain the measured current. Compared to our usual efficiency of $10^{-6}-10^{-8}$ electrons per photon via back-illumination, the rate seems extremely high. Furthermore, the laser is operated at a photon energy of $4.8\,$eV, for which we try to optimize the cathode; hydrogen, on the other hand, features excitations of sufficient energy only above $8\,$eV. This suggests that either the lower energy of our laser, or the back-illumination decreases the efficiency substantially.\\
It seems likely that we did not model the secondary emission by positive ions/neutral molecules correctly. Two factors might come into play, which are non-trivial to implement: the transversal diffusion, which might be non-negligible in hydrogen at low pressures, could "shift" the ion current radially away from the photo-cathode. This might decrease the strength of the secondary emission over time. Secondly, if Phelps assumption of excitation via neutral hydrogen, which is continuously produced by $H_{3}^{+}$ breakdown, is correct, the model is probably too simple. Instead of releasing an electron with a certain probability per arriving positive ion, the hydrogen radicals should be simulated. 



\printbibliography

\end{document}